\documentclass[sigconf]{acmart}

\AtBeginDocument{%
  \providecommand\BibTeX{{%
    \normalfont B\kern-0.5em{\scshape i\kern-0.25em b}\kern-0.8em\TeX}}}
    
\copyrightyear{2022}
\acmYear{2022}
\setcopyright{acmcopyright}
\acmConference[GECCO '22]{Genetic and Evolutionary Computation Conference}{July 9--13, 2022}{Boston, MA, USA}
\acmBooktitle{Genetic and Evolutionary Computation Conference (GECCO '22), July 9--13, 2022, Boston, MA, USA}
\acmPrice{15.00}
\acmDOI{10.1145/3512290.3528861}
\acmISBN{978-1-4503-9237-2/22/07}

\begin{document}

\title{Evolving Programmable Computational Metamaterials}

\author{Atoosa Parsa}
\affiliation{%
  \institution{University of Vermont}
  \city{Burlington}
  \state{Vermont}
  \country{USA}
}
\email{atoosa.parsa@uvm.edu}

\author{Dong Wang}
\affiliation{%
  \institution{Yale University}
  \city{New Haven}
  \state{Connecticut}
  \country{USA}
}
\email{dong.wang@yale.edu}

\author{Corey S. O'Hern}
\affiliation{%
  \institution{Yale University}
  \city{New Haven}
  \state{Connecticut}
  \country{USA}
}
\email{corey.ohern@yale.edu}

\author{Mark D. Shattuck}
\affiliation{%
  \institution{City College of New York} 
  \city{New York}
  \state{New York}
  \country{USA}
}
\email{shattuck@ccny.cuny.edu}

\author{Rebecca Kramer-Bottiglio}
\affiliation{%
  \institution{Yale University}
  \city{New Haven}
  \state{Connecticut}
  \country{USA}
}
\email{rebecca.kramer@yale.edu}

\author{Josh Bongard}
\affiliation{%
  \institution{University of Vermont}
  \city{Burlington}
  \state{Vermont}
  \country{USA}
}
\email{josh.bongard@uvm.edu}

\renewcommand{\shortauthors}{Parsa et al.}

\begin{abstract}
Digital signal processors are widely used in today's computers to perform advanced computational tasks. But, the selection of digital electronics as the physical substrate for computation a hundred years ago was influenced more by technological limitations than substrate appropriateness. In recent decades, advances in chemical, physical and material sciences have provided new options. Granular metamaterials are one such promising target for realizing mechanical computing devices. However, their high-dimensional design space and the unintuitive relationship between microstructure and desired macroscale behavior makes the inverse design problem formidable. In this paper, we use multiobjective evolutionary optimization to solve this inverse problem: we demonstrate the design of basic logic gates embedded in a granular metamaterial, and that the designed material can be ``reprogrammed'' via frequency modulation. As metamaterial design advances, more computationally dense materials may be evolved, amenable to reprogramming  by increasingly sophisticated programming languages written in the frequency domain.


\end{abstract}

\ccsdesc[500]{Computing methodologies~Randomized search}
\ccsdesc[300]{Theory of computation~Models of computation}
\ccsdesc[100]{Hardware~Emerging technologies}

\keywords{Granular Metamaterials, Mechanical Computing, Logic Gates, Inverse Design Problem, Multiobjective Optimization}

\begin{teaserfigure}
  \includegraphics[width=\textwidth]{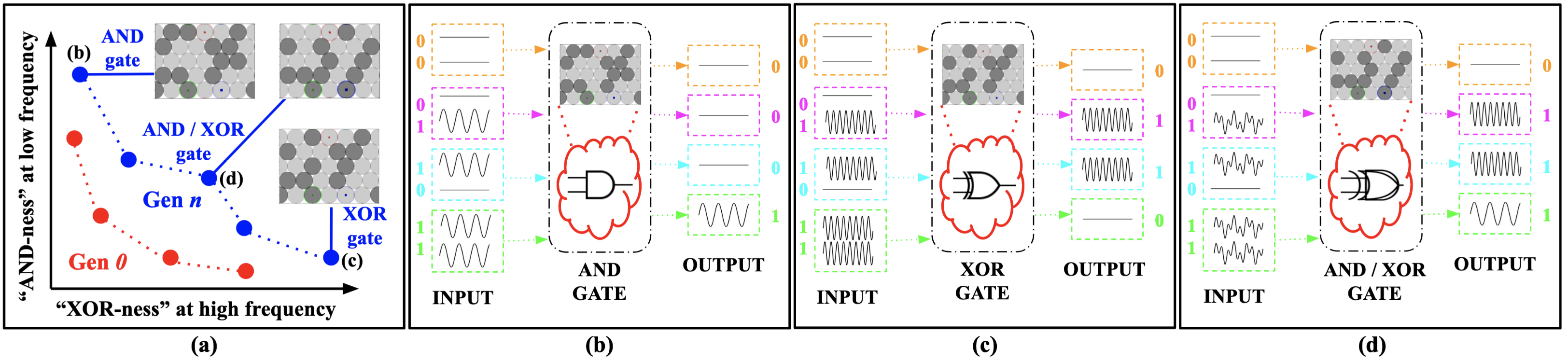}
  \caption{Overview. (a): Multiobjective optimization (MOO) evolves increasing amounts of computational potential into granular metamaterials. Each individual in the population is a granular metamaterial composed of two particle types. Different configurations of particles confer different material behaviours. In the approach reported here, vibrations are supplied as input and vibration (if any) is recorded as output. Materials have been found that act as an AND gate (b) or an XOR gate (c). We report here how MOO can discover a single material (d) that, at one frequency acts as an AND gate, and at another frequency acts as an XOR gate. Thus, the superposition of input waves supplied to the system will result in emergent behaviors other than what the material was originally designed to perform. This suggest future materials amenable to reprogramming using increasingly sophisticated programming languages expressed in the frequency domain.} 
  \label{fig1}
\end{teaserfigure}

\maketitle

\section{Introduction}

Metamaterials are an emerging class of engineered composite materials that exhibit properties different from their constituent materials and behave in ways not observed in nature \cite{kadic20193d}. In the same way that metamaterials exhibit non-intuitive and exotic material properties, computational metamaterials have potential to perform computation in useful ways as well \cite{silva2014performing, zangeneh2021analogue}. 
Here, we explore various ways to evolve a particular class of metamaterial---granular metamaterials---to increase its computational capabilities.
Granular metamaterials are metamaterials consisting of discrete particles that exhibit increased plasticity compared to continuous metamaterials because they can be dynamically reconfigured: particle properties can change in response to external stimuli \cite{wu2019active}. In this paper, we propose using granular metamaterials as a physical substrate to perform mechanical computation.

Considering logic gates as the basic computational blocks upon which more complex units can be built, we demonstrate here the design of metamaterials that can act as basic acoustic logic gates: low or high vibration frequencies or amplitudes can serve as zeros or ones. There are several advantages of this type of computation compared to digital computation where logic gates are built from electrical transistors. First, by moving to a mechanical substrate we can avoid analogue to digital conversion, sidestepping all of the limitations of abstract representations and discretizations necessary for a digital computing system \cite{zangeneh2021analogue}. Second, by supplying vibrations composed of multiple frequencies, multiple computations may be performed simultaneously using the same patch of material, suggesting increasingly computationally dense materials may be evolved in future. Third, acoustically driven computational metamaterials could serve as useful building blocks for more complex machines such as robots: sensing, control and actuation could all respond to and/or produce vibration, rendering these otherwise separate robot components as just different regions within a single-material robot. Finally, by exploiting the natural dynamics of the metamaterial, energy efficient computation and higher robustness and stability may be achieved. Moreover, because computational components do not need to be physically separate modules (as we show below), this raises the possibility of bottom-up design of computational architecture where the exact form of computation is not predetermined \cite{nakajima2020physical}.

In recent years there has been some research on embedding mechanical computation into material. In \cite{serra2019turing}, a universal logic gate is implemented as a nonlinear mass-spring-damper model. In \cite{raney2016stable}, a soft bistable building block is designed and used in the implementation of soft mechanical diodes and logic gates. \cite{li2014granular} and \cite{bilal2017bistable} present examples of acoustic gate design in a 1D chain of elastic particles. Computational metamaterials have been introduced to perform specialized computing tasks such as integration, differentiation and convolution \cite{zangeneh2021analogue}. Perhaps closest to our work is \cite{wu2019active}, in which the authors show the potential of a hand-designed 2D granular system to act as an acoustic switch modulated by the packing pressure to switch between on and off states. Despite these recent advances, in none of the aforementioned works is the computational unit automatically optimized to perform computation, let alone how best to densely pack computation in new ways into materials is explored. Instead, computational building blocks are hand-designed based on the intuitions of a human designer. In this paper, we propose using evolutionary algorithms to automatically optimize the material properties for dense computation. Moreover, the mentioned works are mostly conducted using continuous rather than granular metamaterials, or in one-dimensional particle chains, while our work is performed with two-dimensional sheets of granular metamaterial. Due to advantages of this type of material for dynamic programmability (as we will show), we envision a wider potential in expanding our work to more complex computational elements, as explained in our conclusions.

Granular metamaterials possess many parameters that affect their behavioral response to stimuli. For example, the particle properties (size, shape, stiffness, mass, and so on) and their placements can affect the eigenfrequencies of the system and consequently the localized propagation or suppression of acoustic waves through the material. With so many design parameters, deciding on the optimal micro-structure to achieve a desired macro-behavior (i.e. a logic gate) is a challenge. Evolutionary algorithms have long been shown capable of automated optimization in such design spaces \cite{miskin2013adapting}, \cite{o2013highly}. Given this, in previous work \cite{ourPaper} we demonstrated that we can evolve the configurations of granular metamaterials such that they act as logic gate. Here, we show that a MOO can densely pack multiple operations into the same piece of material: at one frequency the material acts as an AND gate; at another frequency it acts as an XOR gate. Finally, we show that combining two different input frequencies can enable the material to simultaneously compute both logical functions. This suggest future materials may be amenable to reprogramming using increasingly sophisticated programming languages expressed in the frequency domain.

\section{Methods}
In order to build a computing system, we need to choose a physical substrate within which to embed the architecture of the computational model \cite{yasuda2021mechanical}. The conventional choice is digital electronic devices. As discussed in the introduction, this choice was heavily influenced by the manufacturing technology of the time: the promise of the semiconductor industry and Gordon Moore's repeatedly validated prediction about the doubling of transistors on integrated circuits every two years \cite{moore1965cramming}. This promise led to the abandonment of the other computational substrates and huge investments in digital electronics. General purpose digital computers in effect won the ``Hardware Lottery'' and enjoyed significant advances during the last 50 years \cite{hooker2021hardware}. However, the physical constraints of miniaturization are now causing Moore's Law to slow significantly \cite{theis2017end}. This has triggered a renewed interest in exploring other possible physical substrates for computation. In addition to photonics \cite{wu2021analog}, DNA computing \cite{amos2002topics} and quantum computing \cite{hey1999quantum}, Metamaterials are one such promising computational substrate.

In this paper, we explore the computational capabilities of granular metamaterials. Granular metamaterials have been studied to a great extent in the material sciences literature \cite{jenett2020discretely, behringer2018physics, kim2019wave}. They are made from many individual grains with possibly different material properties. They can exhibit complex interesting behaviors but there are (simplified) computational models that can accurately predict them \cite{schreck2010computational}. The basic principle that we use in this work is the ability of a granular assembly to have extinguishable responses to mechanical vibrations with different frequencies. In the next section, we expand this idea and formalize the problem setup.

\subsection{Embedding Computation}
\begin{figure}
\centering
\includegraphics[scale=0.45]{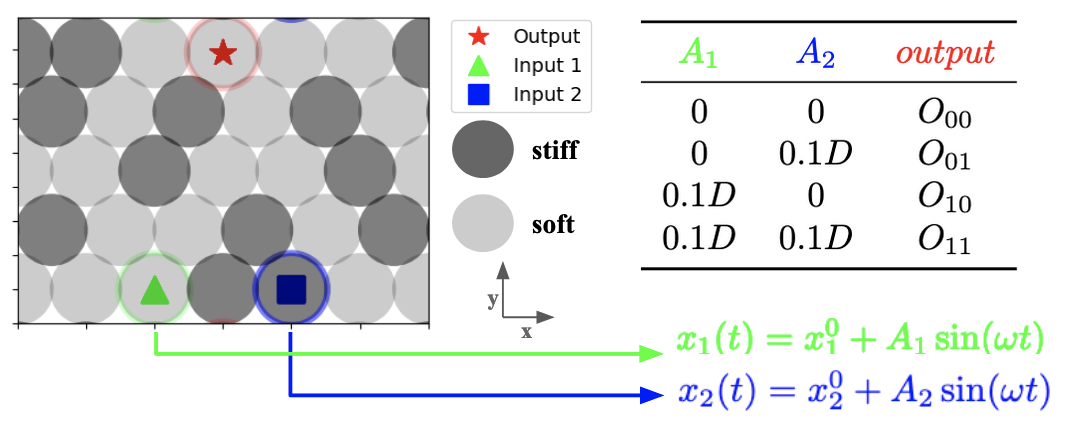}
\caption{The system setup. A logic gate with two inputs (green and blue particles) and one output (red particle) embedded in a 2D granular assembly with two types of particles (indicated with different shades of grey). $A_1$ and $A_2$ are the amplitudes of oscillations applied in $x$ direction to the input ports. $\omega$ is the input frequency. The truth table is shown on the right: $D$ represents the diameter of particles. $O_{ij} (i,j \in {0, 1})$ is the magnitude of the vibrations in the output.}\label{fig2}
\end{figure}

In a typical electronic system, the inputs and outputs are electrical signals. Here, we evolve an analogous mechanical system where inputs and outputs are acoustic signals. Our system is a two dimensional assembly of two types of circular particles placed on a hexagonal lattice. The setup is shown in Fig.~\ref{fig2}. As we mentioned before, such granular system can be widely tuned by changing the particles' properties to achieve different responses \cite{jenett2020discretely}. The response is dependent on a large number of variables, such as particle arrangements, mass, modulus, shape, interactions between particles, and boundary conditions of the system. These properties can affect the frequency spectrum of the material, normal modes of the system, and consequently position of the band gap: a contiguous region of the input frequency spectrum muffled by the material.

Our goal is to regulate the intrinsic material properties to obtain a metamaterial with a desired vibrational response. We are interested in embedding computation in the granular metamaterial and the first step towards this goal is to see if we can design the basic logic gates. In this system, the inputs and outputs are acoustic waves. So we choose two particles on one side of the material as the input ports (the particles with green and blue markers in Fig. \ref{fig2}) and one particle on the other side as the output port (the particle with the red marker). The input signal is a sinusoidal wave with amplitude $A_i$ and frequency $\omega$ in $x$ direction. This signal is applied to a particle, causing it to move from its initial position ($x_i^0$). We can look at the amplitude of the displacement signal in time ($x_i(t)$) as the abstract bit representation: small or no displacement means a \textbf{$`0`$} and high magnitude of displacement means a \textbf{$`1`$}. Going back to our analogy for the input signal, applying a sinusoidal wave with amplitude zero ($A_i=0$) to an input port means a $`0`$ bit at that port of the logic gate. In our simulations, we fixed the non-zero amplitude to $1 \times 10^{-2}$ (which is $10\%$ of the diameter of a particle.). The frequency of the applied signal ($\omega$) is also a fix value, chosen based on the frequency spectrum of a regular configuration which will be discussed in the next section. The truth table in Fig.~\ref{fig1} shows our bit representation. The output $O_{ij} (i,j \in {0, 1})$ is the amplitude of the displacement of the particle at the output port.

In each of the four cases shown in the truth table in Fig.~\ref{fig1}, the gain of the system is defined as the amplitude of the fast Fourier transform ($\hat{f}$) at the driving frequency ($\omega$) in the output, divided by the sum of the amplitudes of the fast Fourier transform at the driving frequency ($\omega$) in the inputs. In other words:

\begin{equation}\label{eq1}
G_{ij}(\omega) = \frac{\hat{f}(O_{ij})}{\hat{f}({\mathrm{in}}_i)+\hat{f}({\mathrm{in}}_j)} \quad i,j \in{0, 1}
\end{equation}

In our experiments, we can measure the gain ($G_{ij}$) for each of the four input cases. Then, in order for the material to act as a logic gate, the relative magnitude of the gain in each case must be consistent with desired functionality of the gate. Note that because it's experimentally impractical to get an absolute zero or one, we are looking at the relative magnitudes of the two cases. For example, for an AND gate when both input ports are driven with a sinusoidal wave ($input = `11`$), we expect to see a high amplitude of oscillation at the output ($output = `1`$) and therefore we expect a high gain ($G_{11}$). But in the other remaining three cases ($input = `00`, `01`$, and $`10`$) we expect a low amplitude of vibration ($output = `0`$) and thus a low gain ($G_{00}, G_{01}$ and $G_{10}$). Based on this formulation, we can devise a metric to measure the similarity of the material's functionality to a desired logic gate. We will show this metric in the next section.

\subsection{The Simulator}
The simulator \footnote{{\href{https://github.com/AtoosaParsa/gecco-2022}{\textbf{https://github.com/AtoosaParsa/gecco-2022}} contains the source code necessary for reproducing our results.}} is a simplified granular system based on the model used in \cite{wu2019active}. It's a two dimensional system, made of frictionless circular disks with a fixed diameter ($D$). The particles can have different material properties such as different masses or different stiffnesses (this is shown as dark/light colors in the figures.). They are placed on a $5$ by $6$ hexagonal lattice and so there is a total number of $30$ particles. The system is periodic in $x$ direction and has a fixed boundary in $y$ direction. Gravity is ignored, so the only forces acting on the particles are the result of a purely repulsive linear spring potential between the disks which can be formalized as a Lennard-Jones potential.

Our system is simulated using discrete element method (DEM). Starting from the initial positions of the particles (placed on a hexagonal lattice), the repulsive forces are calculated based on the distance between the overlapping particles. Next, the accelerations, velocities and positions are updated using the Verlet integration. But there is a pre-processing step to make sure that the system is at an equilibrium at $t=0$ (meaning that the sum of total forces between particles is \textit{(near)} zero) and thus the particle packing is statistically stable. This is done by calculating the total force and updating the initial positions of the particles using the steepest-descent method. Here, we use Fast Inertial Relaxation Engine (FIRE) to reduce the processing load.

Table~\ref{tab:tab1} includes the main simulation parameters and their assigned values in our experiments. $N_t$ is total simulation time.

\begin{table} [h]
  \caption{Simulator Parameters}
  \label{tab:tab1}
  \begin{tabular}{cc}
    \toprule
     Parameter & Value\\
    \midrule
     Particle Diameter & $0.1$\\
     Particle Mass & $1$\\
     Stiffness Ratio & $10$\\
     Packing Fraction & $0.91$\\
     $N_t$ & $1e4$\\
  \bottomrule
\end{tabular}
\end{table}

As was mentioned briefly before, one of the important properties of a granular assembly is the existence of gaps in their vibrational density of states \cite{deymier2013acoustic}. To find the band gap (the biggest gap), the mass-weighted dynamical matrix is calculated using the Hessian of the total potential energy. The eigenvalues of this matrix are the eigenfrequencies of the system. We can plot the eigenfrequency spectrum by sorting the frequencies in an increasing order, then we see the gaps in the spectrum, where We call the biggest gap the \emph{band gap}. When this granular system is excited at a specific frequency, depending on where that frequency falls within the spectrum, the acoustic signal will propagate or will be filtered. This is the underlying process in the granular metamaterial that enables it to act as a logic gate.

\subsection{Performance Measures}
The goal of the optimization is to find a configuration of particles that will act as an AND gate at one frequency ($\omega_1$) and an XOR at another frequency ($\omega_2$). Thus, we need to define metrics for each of these two cases to measure the amount of ``AND-ness'' and ``XOR-ness'' in a candidate solution. We use the same metrics that we defined in our previous work \cite{ourPaper}. As we discussed at the start of this section, we can measure the gain of the system ($G_{ij}(\omega)$ in equation \ref{eq1}) for each of the four possible input cases ($`00`, `01`, `10`, `11`$). $G_{00}$ is trivial: if the input is $`00`$, meaning that no vibration is applied to either of the input particles, the output will remain $`0`$ as well. But for the remaining three cases, we have specific expectations in a logic gate. In an AND gate, we only want to have a significant gain when both of the inputs are activated at $\omega_1$ frequency (high $G_{11}(\omega_1)$ is desired). In order to achieve this goal, one option is to define a single fitness function as follows:

\begin{equation} \label{eq2}
    f_{\textrm{``AND-ness''}}=\frac{G_{11}({\omega_1})}{(G_{10}({\omega_1})+G_{01}({\omega_1}))/2}
\end{equation}

On the other hand, to have an XOR gate at frequency $\omega_2$, we expect to have a significant vibration at the output, if only one of the input ports is being driven by a sinusoidal displacement ($`01`$ and $`10`$ cases). So the fitness in this case can be defined as follows:

\begin{equation} \label{eq3}
    f_{\textrm{``XOR-ness''}}=\frac{(G_{10}({\omega_2})+G_{01}({\omega_2}))/2}{G_{11}({\omega_2})}
\end{equation}

Our goal is to find a configuration of particles that maximizes both of these objectives, so we can formulate the problem as a multiobjective optimization problem. We present the details of the optimization in the next section.

\subsection{Evolutionary Search}
We employ a standard multiobjective optimization algorithm, Nondominated Sorting Genetic Algorithm II (NSGA-II) \cite{deb2002fast}. The hyperparameters used in the simulations are shown in Table~\ref{tab:params} below:

\begin{table} [h]
  \caption{Optimization Parameters}
  \label{tab:params}
  \begin{tabular}{cc}
    \toprule
     Parameter & Value\\
    \midrule
     Population Size & $50$\\
     Generations & $250$\\
     Runs & $5$\\
     Mutation Probability & $0.8$\\
     Crossover Probability & $0.2$\\
     Bit-flip Probability & $0.05$\\
  \bottomrule
\end{tabular}
\end{table}

The individuals are granular assemblies made of two types of particles (soft and stiff particles). We use a direct encoding, so the genome is a binary string of length $30$ (the lattice is $5 \times 6$). In this binary encoding, $`1`$ means a stiff particle (dark grey particles in the figures) and $`0`$ means a soft particle (light grey particles in the figures). In order to add variation to the population, both mutation and crossover operators are implemented. The mutation operator is a bit-flip operator with probability of $0.05$. The crossover is a single-point crossover operator. We implement a $(\lambda+\mu)$ evolutionary algorithm where both $\lambda$ and $\mu$ are equal to the population size. Selection of the Pareto nondominated front is based on the Generalized Reduced Run-Time Complexity Non-Dominated Sorting algorithm presented in \cite{fortin2013generalizing}. We perform $5$ independent runs, each with a different random initial population.

All of our code are written in Python. We use the DEAP library \cite{DEAP_JMLR2012} for the evolutionary optimization. SCOOP \cite{SCOOP_XSEDE2014} is used to parallelize the code to speed up the population evaluation. All of the experiments ran on our computing cluster.

\section{Results}
In this section, we first present our experimental setup using the methods and metrics developed in the previous section. Then, we show the optimization results and take a closer look at the Pareto optimal solutions by investigating their functionality as logic gates.

\subsection{Experimental Setup}
Our goal is to find a granular assembly made of two types of particles that can act as an AND gate at $\omega_1=7$ and as an XOR gate at $\omega_2=10$. These frequencies have been chosen arbitrarily and by looking at the frequency spectrum of random configurations. We formalized the problem as an optimization problem with two objectives (equations \ref{eq2} and \ref{eq3}) and proposed to use the NSGA-II algorithm to solve the problem. The simulator and optimization parameters were also presented in the previous section.

\subsection{Evolved Solutions}
Fig.~\ref{fig3} shows the evolutionary progress. Panel (a) of this figure plots the evolution of the first fitness (``AND-ness'' (Eqn.~\ref{eq2})) on top and the second fitness (``XOR-ness'' (Eqn.~\ref{eq3})) on the bottom. Solid blue line in these two plots shows the average fitness of the population, averaged over $5$ independent evolutionary runs, along with its standard deviation in light blue. We see that the average fitness fluctuates during the evolution and even decreases at a few points. Since the problem is a multiobjective optimization, sometimes new individuals are added to the population that might have lower values in one fitness but higher values in the other. This can cause a decrease in the the average fitness of the whole population. 

In order to show the effectiveness of the evolutionary search, we also performed a random search for each of the two objectives: one for the AND-ness metric and the other for the XOR-ness metric. Since each one of the $30$ particles can either be soft or stiff, there are a total number of $2^{30}=1073741824$ possible configurations. In each experiment, we drew $5000$ random configurations from a uniform distribution and evaluated them based on the objective. Fig.~\ref{fig3}c shows the histograms of these two random distributions. For comparison, we've also plotted the best and worst found solutions in each figure. The histograms show how rugged the fitness landscape is: for the AND-ness metric, the histogram spreads from $0.026$ to $7.88$ with a mean value around $0.760$ and for the XOR-ness, it spreads from $0.21$ to $60.24$ with a mean at $1.622$.

Fig.~\ref{fig3}b shows the Pareto front at three different stages during the evolution for one of the $5$ independent evolutionary runs. We can see that at the start of the optimization (generation $0$), individuals in the population have objective values close to the mean value of a randomly drawn solution. As evolution proceeds, the population progresses towards higher fitness values and also spreads across the axis to the extreme solutions where one objective function (AND-ness/XOR-ness) dominates the other one (see red markers showing the Pareto front at generation $250$). The magenta square on this plot shows one of the solutions in the middle of the Pareto front which has moderately high fitness value in both objectives. the plot on the bottom of this panel shows the chosen particle configuration along with its fitness values.

\begin{figure}[h]
    \centering
    \includegraphics[scale=0.30]{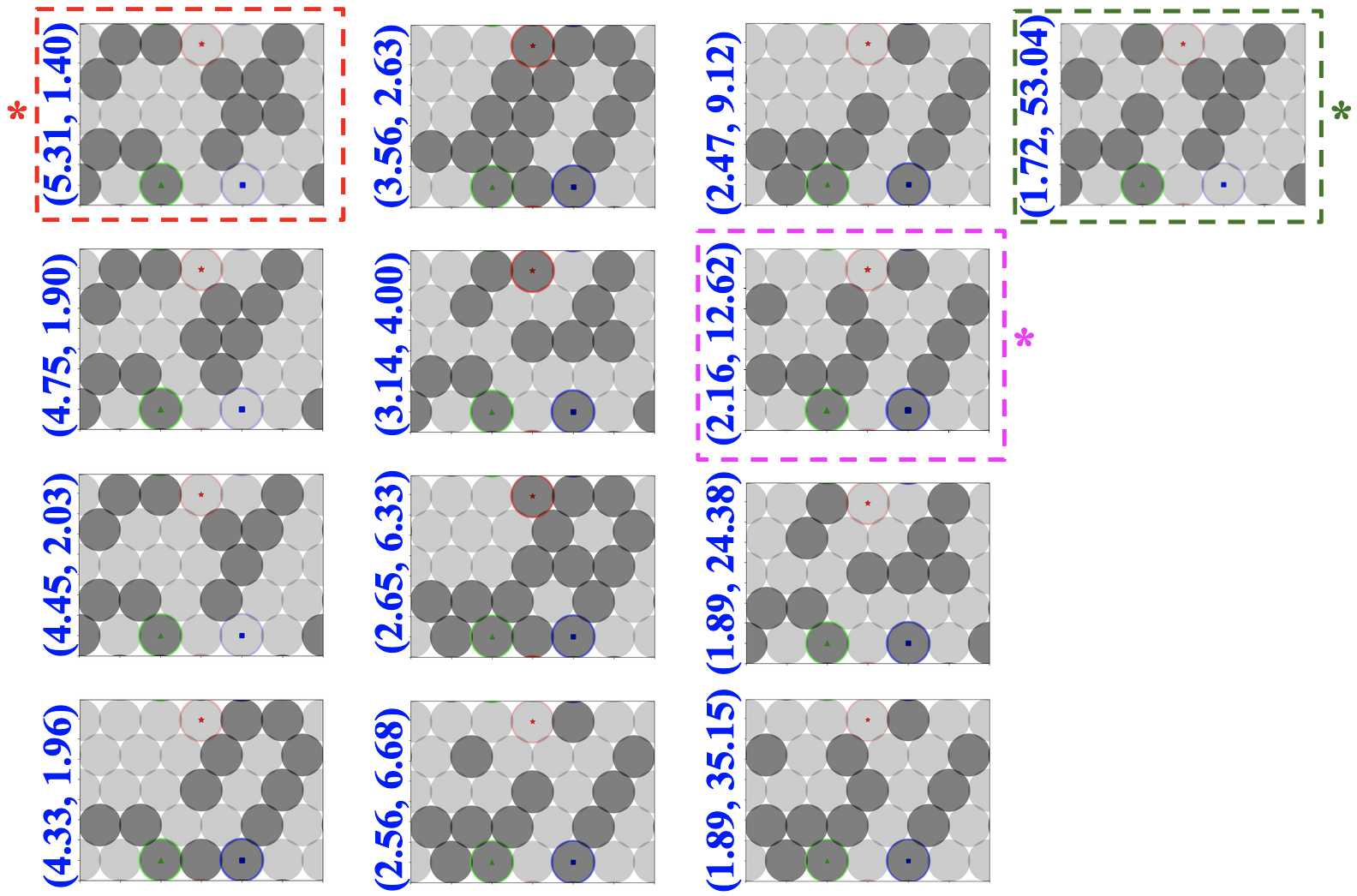}
    \caption{Configurations in the non-dominated Pareto front at the last generation of the evolutionary process. The magenta star shows the chosen solution, the green and red stars show the extreme solutions with highest AND-ness or highest XOR-ness. The numbers beside each configuration show the fitness values as (AND-ness, XOR-ness). These are the candidate solution of our optimization problem: can you find any specific patterns common in the particle placements shown in this figure?}
    \label{fig4}
\end{figure}

All the solutions of the last generation of the evolutionary search are Pareto-optimal. These solutions are shown in Fig.~\ref{fig4}. It's interesting to investigate these configurations visually, to see if one can find a regular pattern or a specific order in the particle placements that correlates with high fitness values. We'll talk about this in more detail in the Discussion section. The magenta star in this figure shows a candidate solution with moderately high fitness values in both objectives. This is the configuration that we chose to investigate further in Fig.~\ref{fig5}.

\begin{figure*}[h]
    \centering
    \includegraphics[scale=0.31]{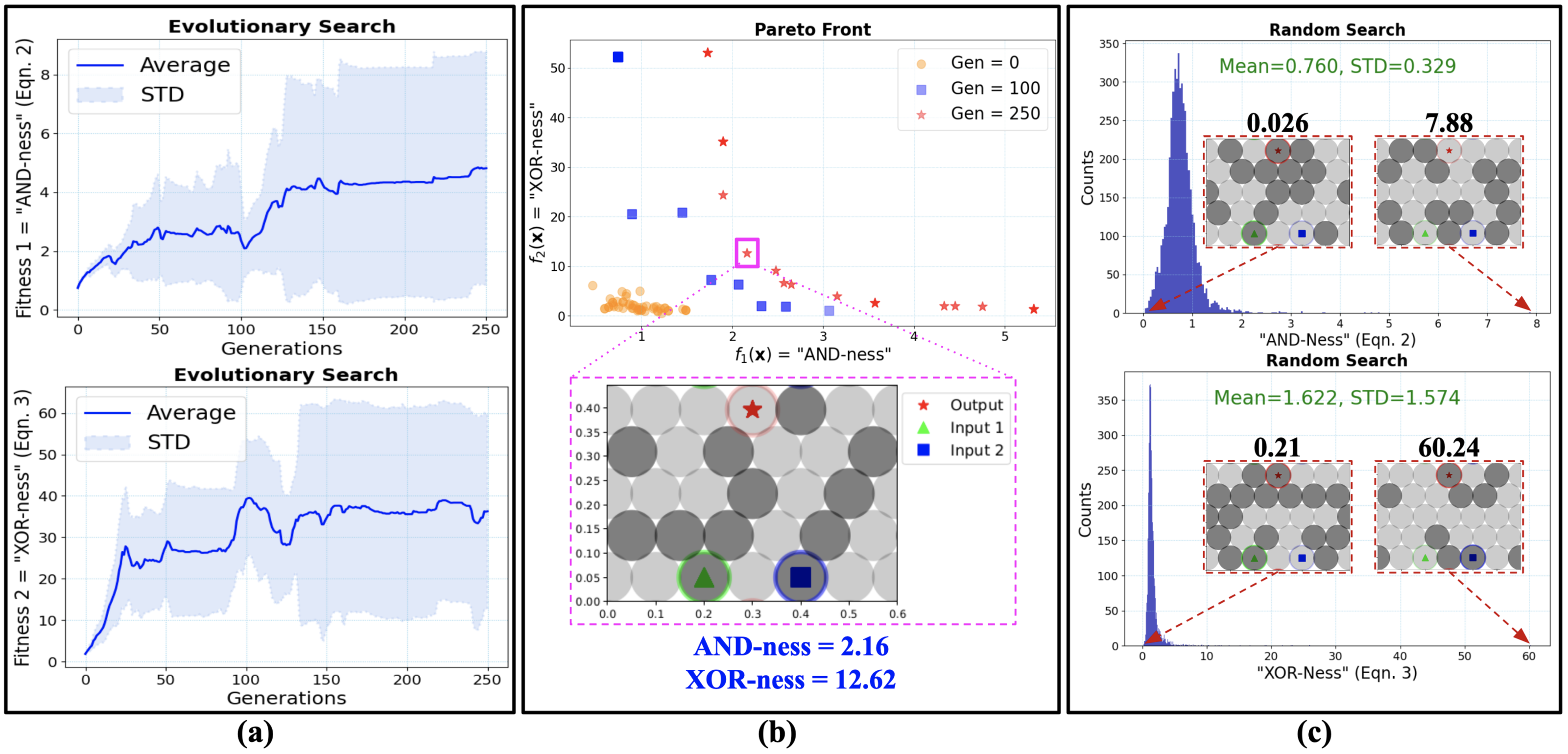}
    \caption{Evolutionary progress. The progress of each of the objective functions during the evolution (averaged over 5 independent runs) is plotted in (a) for AND-ness (top) and XOR-ness (bottom). (b) shows the Pareto front of one of the evolutionary runs at three different stages during the optimization. The magenta square in this plot indicates the chosen solution from the final population and its corresponding particle configuration. (c) shows histograms of randomly drawn configurations. They demonstrate that a random configuration will on average have an AND-ness of $0.760$ and XOR-ness of $1.622$.}
    \label{fig3}
\end{figure*}

In order to examine the performance of the evolved solution, we tested the response of the configuration when different vibrations are applied to the input ports. Fig.~\ref{fig5}a shows the functionality of the solution as an AND gate at $\omega_1=7$. Fig.~\ref{fig5}b shows its function as an XOR gate at $\omega_1=10$. For each gate, we supply the four different input signals: $`00`$, $`01`$, $`10`$ and $`11`$. For the AND gate, we see a high amplitude of oscillation when both of the input ports are activated at $\omega_1=7$ (see $(v)$ in panel (a)). The response is plotted both in frequency and in time space. In the XOR case in panel (b), only when one of the inputs is activated we see a high magnitude of displacement at the output. The plots confirm the multifunctionality of the designed granular assembly at different frequencies. In other words, we can ``program'' the material by modulating the frequency of the input signal. This means that without needing to reconfigure the particles or change any material properties, we can observe different functionalities from the material.

\begin{figure*} [h!]
    \centering
    \includegraphics[scale=0.49]{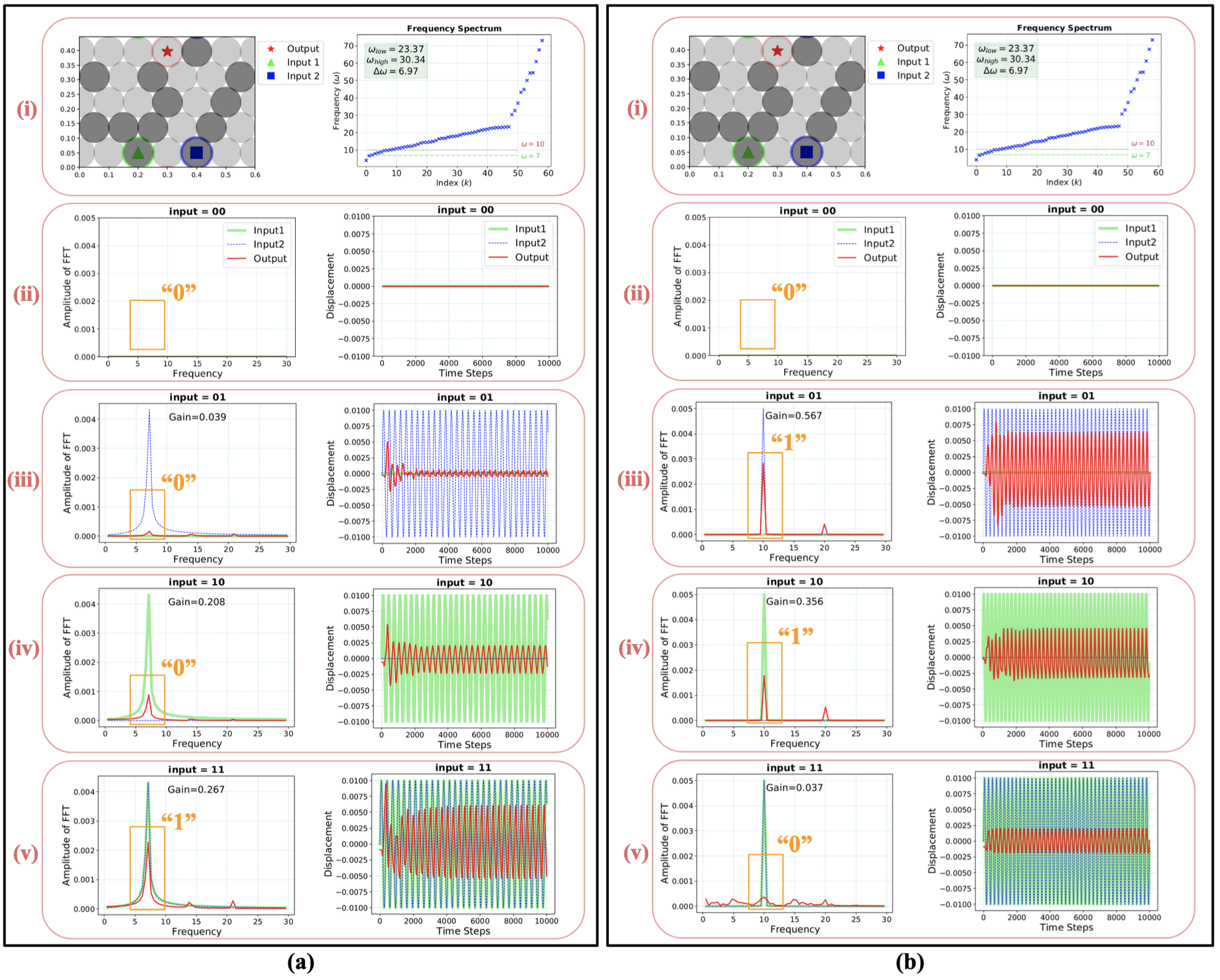}
    \caption{Evaluating the functionality of the candidate solution. (a) examines its performance as an AND gate at $\omega=7$. Panel $(i)$ shows the particle configuration on the left and the frequency spectrum on the right. The horizontal lines mark the excitation frequencies at $\omega_1$ and $\omega_2$. Panels $(ii)$ through $(v)$ show the response of the system in frequency and time space for the four input cases. (b) examines the performance of the candidate configuration as an XOR gate at $\omega=10$.}
    \label{fig5}
\end{figure*}

\section{Discussion and Conclusion}
In our previous paper \cite{ourPaper}, we introduced the possibility of performing computation in a granular substrate. We showed that we can design a granular assembly made of two types of particles which can act as an AND gate or an XOR gate depending on the frequency of the input vibrations. But one of the reasons that mechanical computing devices were abandoned at the outset of the computer age and overshadowed by their digital electronic counterparts was the miniaturization capability of electronic transistors and the promise of the compact computational power predicted by Moore's Law. To address this concern in the application of granular metamaterials as a new generation of computational devices, we decided to investigate how much computational power can be packed into one granular configuration. Because of the unique properties of a granular metamaterial under vibrations with different frequencies, we decided to work on modulating the input frequency as a way to program the material to perform different computational tasks. In the Results section, we demonstrated the success of our approach in finding particle configurations that can exhibit two different functionalities without any changes to the material after fabrication.

\begin{figure*}[h!]
    \centering
    \includegraphics[scale=0.41]{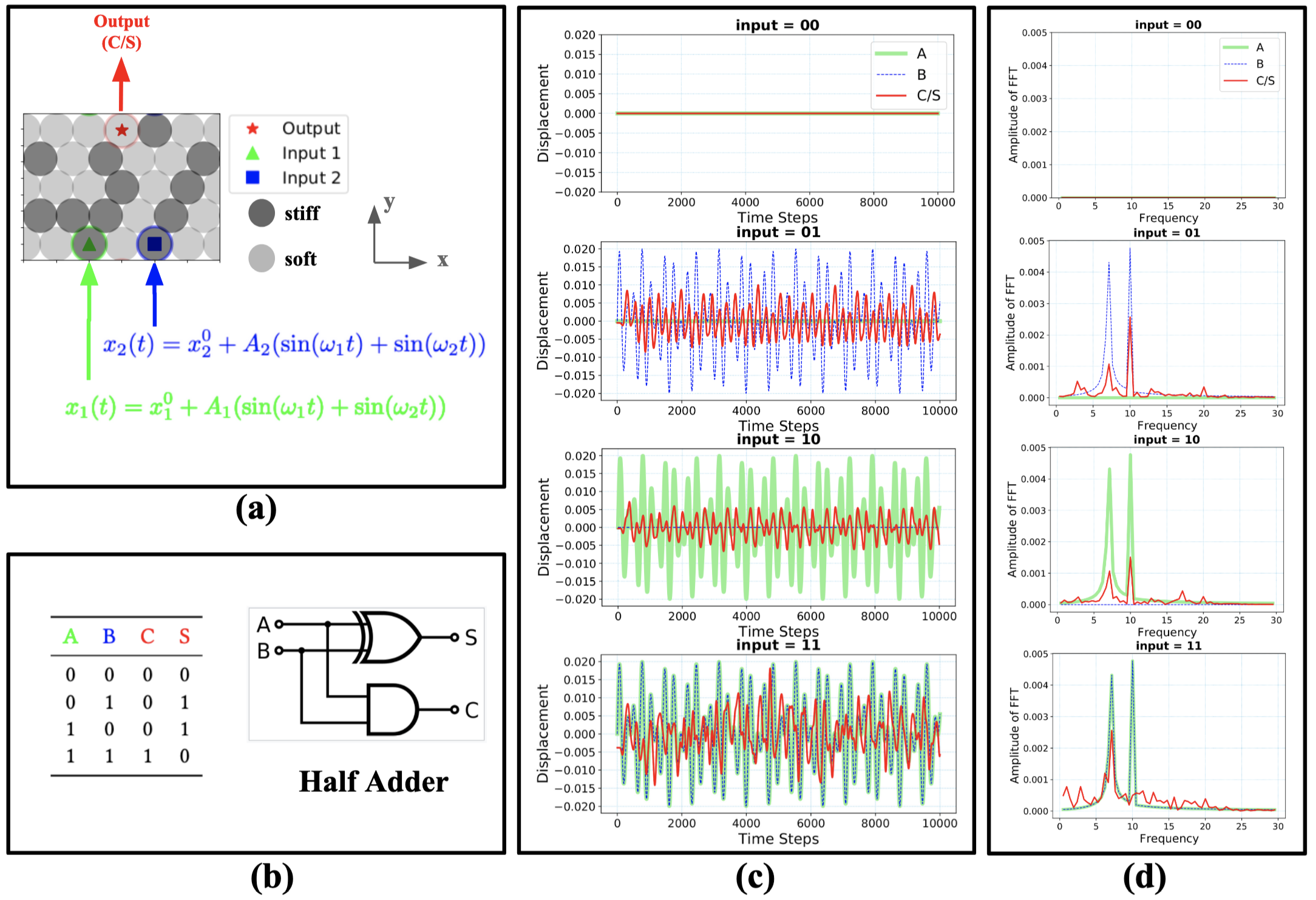}
    \caption{Programming the granular metamaterial to act as a half adder.  (a) shows the optimal configuration and the input signals. The material was designed to act as an AND gate at $\omega=7$ and as an XOR gate at $\omega=10$. The inputs are the sum of two sinusoidal waves with these two frequencies. (b) reports the logic diagram of a half adder along with its truth table. (c) shows the time-series response of the programmable granular metamaterial in the four possible cases: sum of two sinusoidal waves with frequencies $\omega_1=7$ and $\omega_2=10$ is applied to none of the input ports ($`00`$), input 1 ($`01`$), input 2 ($`10`$) or both ($`11`$). (d) demonstrates the response of the system in the frequency spectrum: the spikes in each signal indicate a $`1`$ at that frequency. The material acts as an AND gate at $\omega_1$, producing the carry signal (C) in the half adder. At $\omega_2$ the material acts as an XOR gate and so the output port produces the sum signal (S) in the half adder.}
    \label{fig6}
\end{figure*}

The input frequency is the key to regulate the functionality of the designed metamaterial and potentially increase its computational density. To explore this potential, we investigated the incorporation of a more complex computational block into metamaterials: a half adder. The half adder is a logical computational block that adds two input bits and generates a carry (C) and a sum (S) signal, with the carry signal representing the overflow to the next digit. The simplest half adder design is made of one AND gate and one XOR gate and is shown in  Fig.~\ref{fig6}b. Since the half adder can be implemented using just AND and XOR gates, it is an interesting test case for our designed metamaterial. As shown in Fig.~\ref{fig6}a, we can change the input vibration to the sum of two sinusoidal waves with two different frequencies ($\omega_1 and \omega_2$). Since the output of the computational block is defined as the magnitude of the oscillation at the excitation frequency (refer to the Methods section) in this system, we will have two (temporal) outputs: one at $\omega_1$ and the other at $\omega_2$. The frequency responses in Fig.~\ref{fig6}d demonstrates this idea: there are two spikes in each plot, corresponding to the system response at two different frequencies: one at $\omega_1=7$ which produces the carry signal (C) and the other at $\omega_2=10$ which produces the sum (S).

One of the important performance metrics in this system is the variation between the magnitude of vibrations in $`0`$ and $`1`$ output cases: the higher variation results in a more distinguishable \textit{on} and \textit{off} states. We notice that in the third plot of Fig.~\ref{fig6}d, when the input is $`10`$, the outputs are expected to be $`0`$ at $C$ and $`1`$ at $S$. But the two spikes at $7$ and $10$ frequencies are almost of the same magnitude and the difference is not significant. One way to fix this issue is to choose different excitation frequencies, meaning that instead of choosing $\omega_1=7$ and $\omega_1=10$ we can have frequencies that are further away from each other in the spectrum. This might reduce the possibility that the two phases of the input signal interrupt each other and thus result in a better multifunctional metamaterial. The other option is to increase the resolution of the granular metamaterial. By adding more particles in the system, the frequency spectrum of the material will expand (because the number of degrees of freedom increases). Therefore we will have more normal modes in the system and have more freedom in designing the material.


This paper is meant to confirm the computational power of a granular metamaterial and demonstrate one possible method to program such systems. We are not trying to imply that they have higher performance with regards to traditional digital devices. Much work remains before a fair comparison of the computational power of such a system with today's general purpose digital computers can be performed as the digital electronics and semiconductor industry have benefited from billion dollar investments and huge research efforts. However, it is our hope that this work might help revive the idea of analogue mechanical computing by highlighting their computational potential.

In the end, we would also like to point out the potential of granular metamaterials. In this work, we showcased an overly simplified model made of frictionless circular particles with just one different material property (stiffness). There are other parameters such as particle shapes, masses, modulus, friction and other inter-particle interactions that can be incorporated in building a more sophisticated model. There is also the possibility of utilizing particles that can change their physical properties dynamically in response to some stimuli such as temperature. Such particles can be used to make a ``reconfigurable'' granular assembly with the ability to shift from one configuration to another without the need to reassemble the whole material. There are many possible ways to reconfigure metamaterials to exhibit more complex behaviour yet to be explored, some of which may allow these materials to assume much computational power.

\section{Acknowledgments}
We would like to acknowledge financial support from National Science Foundation under the DMREF program (award number: 2118810). We also acknowledge computation provided by the Vermont Advanced Computing Core.

\bibliographystyle{ACM-Reference-Format}
\bibliography{main}


\begin{thebibliography}{28}


\ifx \showCODEN    \undefined \def \showCODEN     #1{\unskip}     \fi
\ifx \showDOI      \undefined \def \showDOI       #1{#1}\fi
\ifx \showISBNx    \undefined \def \showISBNx     #1{\unskip}     \fi
\ifx \showISBNxiii \undefined \def \showISBNxiii  #1{\unskip}     \fi
\ifx \showISSN     \undefined \def \showISSN      #1{\unskip}     \fi
\ifx \showLCCN     \undefined \def \showLCCN      #1{\unskip}     \fi
\ifx \shownote     \undefined \def \shownote      #1{#1}          \fi
\ifx \showarticletitle \undefined \def \showarticletitle #1{#1}   \fi
\ifx \showURL      \undefined \def \showURL       {\relax}        \fi
\providecommand\bibfield[2]{#2}
\providecommand\bibinfo[2]{#2}
\providecommand\natexlab[1]{#1}
\providecommand\showeprint[2][]{arXiv:#2}

\bibitem[\protect\citeauthoryear{Amos, P{\u{a}}un, Rozenberg, and Salomaa}{Amos
  et~al\mbox{.}}{2002}]%
        {amos2002topics}
\bibfield{author}{\bibinfo{person}{Martyn Amos}, \bibinfo{person}{Gheorghe
  P{\u{a}}un}, \bibinfo{person}{Grzegorz Rozenberg}, {and}
  \bibinfo{person}{Arto Salomaa}.} \bibinfo{year}{2002}\natexlab{}.
\newblock \showarticletitle{Topics in the theory of DNA computing}.
\newblock \bibinfo{journal}{\emph{Theoretical computer science}}
  \bibinfo{volume}{287}, \bibinfo{number}{1} (\bibinfo{year}{2002}),
  \bibinfo{pages}{3--38}.
\newblock


\bibitem[\protect\citeauthoryear{Behringer and Chakraborty}{Behringer and
  Chakraborty}{2018}]%
        {behringer2018physics}
\bibfield{author}{\bibinfo{person}{Robert~P Behringer} {and}
  \bibinfo{person}{Bulbul Chakraborty}.} \bibinfo{year}{2018}\natexlab{}.
\newblock \showarticletitle{The physics of jamming for granular materials: a
  review}.
\newblock \bibinfo{journal}{\emph{Reports on Progress in Physics}}
  \bibinfo{volume}{82}, \bibinfo{number}{1} (\bibinfo{year}{2018}),
  \bibinfo{pages}{012601}.
\newblock


\bibitem[\protect\citeauthoryear{Bilal, Foehr, and Daraio}{Bilal
  et~al\mbox{.}}{2017}]%
        {bilal2017bistable}
\bibfield{author}{\bibinfo{person}{Osama~R Bilal}, \bibinfo{person}{Andr{\'e}
  Foehr}, {and} \bibinfo{person}{Chiara Daraio}.}
  \bibinfo{year}{2017}\natexlab{}.
\newblock \showarticletitle{Bistable metamaterial for switching and cascading
  elastic vibrations}.
\newblock \bibinfo{journal}{\emph{Proceedings of the National Academy of
  Sciences}} \bibinfo{volume}{114}, \bibinfo{number}{18}
  (\bibinfo{year}{2017}), \bibinfo{pages}{4603--4606}.
\newblock


\bibitem[\protect\citeauthoryear{Deb, Pratap, Agarwal, and Meyarivan}{Deb
  et~al\mbox{.}}{2002}]%
        {deb2002fast}
\bibfield{author}{\bibinfo{person}{Kalyanmoy Deb}, \bibinfo{person}{Amrit
  Pratap}, \bibinfo{person}{Sameer Agarwal}, {and} \bibinfo{person}{TAMT
  Meyarivan}.} \bibinfo{year}{2002}\natexlab{}.
\newblock \showarticletitle{A fast and elitist multiobjective genetic
  algorithm: NSGA-II}.
\newblock \bibinfo{journal}{\emph{IEEE transactions on evolutionary
  computation}} \bibinfo{volume}{6}, \bibinfo{number}{2}
  (\bibinfo{year}{2002}), \bibinfo{pages}{182--197}.
\newblock


\bibitem[\protect\citeauthoryear{Deymier}{Deymier}{2013}]%
        {deymier2013acoustic}
\bibfield{author}{\bibinfo{person}{Pierre~A Deymier}.}
  \bibinfo{year}{2013}\natexlab{}.
\newblock \bibinfo{booktitle}{\emph{Acoustic metamaterials and phononic
  crystals}}. Vol.~\bibinfo{volume}{173}.
\newblock \bibinfo{publisher}{Springer Science \& Business Media}.
\newblock


\bibitem[\protect\citeauthoryear{Fortin, {De Rainville}, Gardner, Parizeau, and
  Gagn\'e}{Fortin et~al\mbox{.}}{2012}]%
        {DEAP_JMLR2012}
\bibfield{author}{\bibinfo{person}{F\'elix-Antoine Fortin},
  \bibinfo{person}{Fran\c{c}ois-Michel {De Rainville}},
  \bibinfo{person}{Marc-Andr\'e Gardner}, \bibinfo{person}{Marc Parizeau},
  {and} \bibinfo{person}{Christian Gagn\'e}.} \bibinfo{year}{2012}\natexlab{}.
\newblock \showarticletitle{{DEAP}: Evolutionary Algorithms Made Easy}.
\newblock \bibinfo{journal}{\emph{Journal of Machine Learning Research}}
  \bibinfo{volume}{13} (\bibinfo{date}{jul} \bibinfo{year}{2012}),
  \bibinfo{pages}{2171--2175}.
\newblock


\bibitem[\protect\citeauthoryear{Fortin, Grenier, and Parizeau}{Fortin
  et~al\mbox{.}}{2013}]%
        {fortin2013generalizing}
\bibfield{author}{\bibinfo{person}{F{\'e}lix-Antoine Fortin},
  \bibinfo{person}{Simon Grenier}, {and} \bibinfo{person}{Marc Parizeau}.}
  \bibinfo{year}{2013}\natexlab{}.
\newblock \showarticletitle{Generalizing the improved run-time complexity
  algorithm for non-dominated sorting}. In
  \bibinfo{booktitle}{\emph{Proceedings of the 15th annual conference on
  Genetic and evolutionary computation}}. \bibinfo{pages}{615--622}.
\newblock


\bibitem[\protect\citeauthoryear{Hey}{Hey}{1999}]%
        {hey1999quantum}
\bibfield{author}{\bibinfo{person}{Tony Hey}.} \bibinfo{year}{1999}\natexlab{}.
\newblock \showarticletitle{Quantum computing: an introduction}.
\newblock \bibinfo{journal}{\emph{Computing \& Control Engineering Journal}}
  \bibinfo{volume}{10}, \bibinfo{number}{3} (\bibinfo{year}{1999}),
  \bibinfo{pages}{105--112}.
\newblock


\bibitem[\protect\citeauthoryear{Hold-Geoffroy, Gagnon, and
  Parizeau}{Hold-Geoffroy et~al\mbox{.}}{2014}]%
        {SCOOP_XSEDE2014}
\bibfield{author}{\bibinfo{person}{Yannick Hold-Geoffroy},
  \bibinfo{person}{Olivier Gagnon}, {and} \bibinfo{person}{Marc Parizeau}.}
  \bibinfo{year}{2014}\natexlab{}.
\newblock \showarticletitle{Once you SCOOP, no need to fork}. In
  \bibinfo{booktitle}{\emph{Proceedings of the 2014 Annual Conference on
  Extreme Science and Engineering Discovery Environment}}. ACM,
  \bibinfo{pages}{60}.
\newblock


\bibitem[\protect\citeauthoryear{Hooker}{Hooker}{2021}]%
        {hooker2021hardware}
\bibfield{author}{\bibinfo{person}{Sara Hooker}.}
  \bibinfo{year}{2021}\natexlab{}.
\newblock \showarticletitle{The hardware lottery}.
\newblock \bibinfo{journal}{\emph{Commun. ACM}} \bibinfo{volume}{64},
  \bibinfo{number}{12} (\bibinfo{year}{2021}), \bibinfo{pages}{58--65}.
\newblock


\bibitem[\protect\citeauthoryear{Jenett, Cameron, Tourlomousis, Rubio, Ochalek,
  and Gershenfeld}{Jenett et~al\mbox{.}}{2020}]%
        {jenett2020discretely}
\bibfield{author}{\bibinfo{person}{Benjamin Jenett},
  \bibinfo{person}{Christopher Cameron}, \bibinfo{person}{Filippos
  Tourlomousis}, \bibinfo{person}{Alfonso~Parra Rubio}, \bibinfo{person}{Megan
  Ochalek}, {and} \bibinfo{person}{Neil Gershenfeld}.}
  \bibinfo{year}{2020}\natexlab{}.
\newblock \showarticletitle{Discretely assembled mechanical metamaterials}.
\newblock \bibinfo{journal}{\emph{Science advances}} \bibinfo{volume}{6},
  \bibinfo{number}{47} (\bibinfo{year}{2020}), \bibinfo{pages}{eabc9943}.
\newblock


\bibitem[\protect\citeauthoryear{Kadic, Milton, van Hecke, and Wegener}{Kadic
  et~al\mbox{.}}{2019}]%
        {kadic20193d}
\bibfield{author}{\bibinfo{person}{Muamer Kadic}, \bibinfo{person}{Graeme~W
  Milton}, \bibinfo{person}{Martin van Hecke}, {and} \bibinfo{person}{Martin
  Wegener}.} \bibinfo{year}{2019}\natexlab{}.
\newblock \showarticletitle{3D metamaterials}.
\newblock \bibinfo{journal}{\emph{Nature Reviews Physics}} \bibinfo{volume}{1},
  \bibinfo{number}{3} (\bibinfo{year}{2019}), \bibinfo{pages}{198--210}.
\newblock


\bibitem[\protect\citeauthoryear{Kim and Yang}{Kim and Yang}{2019}]%
        {kim2019wave}
\bibfield{author}{\bibinfo{person}{Eunho Kim} {and} \bibinfo{person}{Jinkyu
  Yang}.} \bibinfo{year}{2019}\natexlab{}.
\newblock \showarticletitle{Wave propagation in granular metamaterials}.
\newblock \bibinfo{journal}{\emph{Functional Composites and Structures}}
  \bibinfo{volume}{1}, \bibinfo{number}{1} (\bibinfo{year}{2019}),
  \bibinfo{pages}{012002}.
\newblock


\bibitem[\protect\citeauthoryear{Li, Anzel, Yang, Kevrekidis, and Daraio}{Li
  et~al\mbox{.}}{2014}]%
        {li2014granular}
\bibfield{author}{\bibinfo{person}{Feng Li}, \bibinfo{person}{Paul Anzel},
  \bibinfo{person}{Jinkyu Yang}, \bibinfo{person}{Panayotis~G Kevrekidis},
  {and} \bibinfo{person}{Chiara Daraio}.} \bibinfo{year}{2014}\natexlab{}.
\newblock \showarticletitle{Granular acoustic switches and logic elements}.
\newblock \bibinfo{journal}{\emph{Nature communications}} \bibinfo{volume}{5},
  \bibinfo{number}{1} (\bibinfo{year}{2014}), \bibinfo{pages}{1--6}.
\newblock


\bibitem[\protect\citeauthoryear{Miskin and Jaeger}{Miskin and Jaeger}{2013}]%
        {miskin2013adapting}
\bibfield{author}{\bibinfo{person}{Marc~Z Miskin} {and}
  \bibinfo{person}{Heinrich~M Jaeger}.} \bibinfo{year}{2013}\natexlab{}.
\newblock \showarticletitle{Adapting granular materials through artificial
  evolution}.
\newblock \bibinfo{journal}{\emph{Nature materials}} \bibinfo{volume}{12},
  \bibinfo{number}{4} (\bibinfo{year}{2013}), \bibinfo{pages}{326--331}.
\newblock


\bibitem[\protect\citeauthoryear{Moore et~al\mbox{.}}{Moore
  et~al\mbox{.}}{1965}]%
        {moore1965cramming}
\bibfield{author}{\bibinfo{person}{Gordon~E Moore} {et~al\mbox{.}}}
  \bibinfo{year}{1965}\natexlab{}.
\newblock \bibinfo{title}{Cramming more components onto integrated circuits}.
\newblock
\newblock


\bibitem[\protect\citeauthoryear{Nakajima}{Nakajima}{2020}]%
        {nakajima2020physical}
\bibfield{author}{\bibinfo{person}{Kohei Nakajima}.}
  \bibinfo{year}{2020}\natexlab{}.
\newblock \showarticletitle{Physical reservoir computing—an introductory
  perspective}.
\newblock \bibinfo{journal}{\emph{Japanese Journal of Applied Physics}}
  \bibinfo{volume}{59}, \bibinfo{number}{6} (\bibinfo{year}{2020}),
  \bibinfo{pages}{060501}.
\newblock


\bibitem[\protect\citeauthoryear{O'Hern and Shattuck}{O'Hern and
  Shattuck}{2013}]%
        {o2013highly}
\bibfield{author}{\bibinfo{person}{Corey~S O'Hern} {and}
  \bibinfo{person}{Mark~D Shattuck}.} \bibinfo{year}{2013}\natexlab{}.
\newblock \showarticletitle{Highly evolved grains}.
\newblock \bibinfo{journal}{\emph{Nature materials}} \bibinfo{volume}{12},
  \bibinfo{number}{4} (\bibinfo{year}{2013}), \bibinfo{pages}{287--288}.
\newblock


\bibitem[\protect\citeauthoryear{Parsa, Wang, O’Hern, Shattuck,
  Kramer-Bottiglio, and Bongard}{Parsa et~al\mbox{.}}{2022}]%
        {ourPaper}
\bibfield{author}{\bibinfo{person}{Atoosa Parsa}, \bibinfo{person}{Dong Wang},
  \bibinfo{person}{Corey~S. O’Hern}, \bibinfo{person}{Mark~D. Shattuck},
  \bibinfo{person}{Rebecca Kramer-Bottiglio}, {and} \bibinfo{person}{Josh
  Bongard}.} \bibinfo{year}{2022}\natexlab{}.
\newblock \showarticletitle{Evolution of Acoustic Logic Gates in Granular
  Metamaterials}. In \bibinfo{booktitle}{\emph{International Conference on the
  Applications of Evolutionary Computation (Part of EvoStar)}}. Springer.
\newblock
\newblock
\shownote{(in press).}


\bibitem[\protect\citeauthoryear{Raney, Nadkarni, Daraio, Kochmann, Lewis, and
  Bertoldi}{Raney et~al\mbox{.}}{2016}]%
        {raney2016stable}
\bibfield{author}{\bibinfo{person}{Jordan~R Raney}, \bibinfo{person}{Neel
  Nadkarni}, \bibinfo{person}{Chiara Daraio}, \bibinfo{person}{Dennis~M
  Kochmann}, \bibinfo{person}{Jennifer~A Lewis}, {and} \bibinfo{person}{Katia
  Bertoldi}.} \bibinfo{year}{2016}\natexlab{}.
\newblock \showarticletitle{Stable propagation of mechanical signals in soft
  media using stored elastic energy}.
\newblock \bibinfo{journal}{\emph{Proceedings of the National Academy of
  Sciences}} \bibinfo{volume}{113}, \bibinfo{number}{35}
  (\bibinfo{year}{2016}), \bibinfo{pages}{9722--9727}.
\newblock


\bibitem[\protect\citeauthoryear{Schreck and O’Hern}{Schreck and
  O’Hern}{2010}]%
        {schreck2010computational}
\bibfield{author}{\bibinfo{person}{CF Schreck} {and} \bibinfo{person}{CS
  O’Hern}.} \bibinfo{year}{2010}\natexlab{}.
\newblock \showarticletitle{Computational methods to study jammed systems}.
\newblock \bibinfo{journal}{\emph{Experimental and computational techniques in
  soft condensed matter physics}} (\bibinfo{year}{2010}),
  \bibinfo{pages}{25--61}.
\newblock


\bibitem[\protect\citeauthoryear{Serra-Garcia}{Serra-Garcia}{2019}]%
        {serra2019turing}
\bibfield{author}{\bibinfo{person}{Marc Serra-Garcia}.}
  \bibinfo{year}{2019}\natexlab{}.
\newblock \showarticletitle{Turing-complete mechanical processor via automated
  nonlinear system design}.
\newblock \bibinfo{journal}{\emph{Physical Review E}} \bibinfo{volume}{100},
  \bibinfo{number}{4} (\bibinfo{year}{2019}), \bibinfo{pages}{042202}.
\newblock


\bibitem[\protect\citeauthoryear{Silva, Monticone, Castaldi, Galdi, Al{\`u},
  and Engheta}{Silva et~al\mbox{.}}{2014}]%
        {silva2014performing}
\bibfield{author}{\bibinfo{person}{Alexandre Silva}, \bibinfo{person}{Francesco
  Monticone}, \bibinfo{person}{Giuseppe Castaldi}, \bibinfo{person}{Vincenzo
  Galdi}, \bibinfo{person}{Andrea Al{\`u}}, {and} \bibinfo{person}{Nader
  Engheta}.} \bibinfo{year}{2014}\natexlab{}.
\newblock \showarticletitle{Performing mathematical operations with
  metamaterials}.
\newblock \bibinfo{journal}{\emph{Science}} \bibinfo{volume}{343},
  \bibinfo{number}{6167} (\bibinfo{year}{2014}), \bibinfo{pages}{160--163}.
\newblock


\bibitem[\protect\citeauthoryear{Theis and Wong}{Theis and Wong}{2017}]%
        {theis2017end}
\bibfield{author}{\bibinfo{person}{Thomas~N Theis} {and}
  \bibinfo{person}{H-S~Philip Wong}.} \bibinfo{year}{2017}\natexlab{}.
\newblock \showarticletitle{The end of moore's law: A new beginning for
  information technology}.
\newblock \bibinfo{journal}{\emph{Computing in Science \& Engineering}}
  \bibinfo{volume}{19}, \bibinfo{number}{2} (\bibinfo{year}{2017}),
  \bibinfo{pages}{41--50}.
\newblock


\bibitem[\protect\citeauthoryear{Wu, Lin, Guo, Liu, Fang, Jiao, and Dai}{Wu
  et~al\mbox{.}}{2021}]%
        {wu2021analog}
\bibfield{author}{\bibinfo{person}{Jiamin Wu}, \bibinfo{person}{Xing Lin},
  \bibinfo{person}{Yuchen Guo}, \bibinfo{person}{Junwei Liu},
  \bibinfo{person}{Lu Fang}, \bibinfo{person}{Shuming Jiao}, {and}
  \bibinfo{person}{Qionghai Dai}.} \bibinfo{year}{2021}\natexlab{}.
\newblock \showarticletitle{Analog Optical Computing for Artificial
  Intelligence}.
\newblock \bibinfo{journal}{\emph{Engineering}} (\bibinfo{year}{2021}).
\newblock


\bibitem[\protect\citeauthoryear{Wu, Cui, Bertrand, Shattuck, and O'Hern}{Wu
  et~al\mbox{.}}{2019}]%
        {wu2019active}
\bibfield{author}{\bibinfo{person}{Qikai Wu}, \bibinfo{person}{Chunyang Cui},
  \bibinfo{person}{Thibault Bertrand}, \bibinfo{person}{Mark~D Shattuck}, {and}
  \bibinfo{person}{Corey~S O'Hern}.} \bibinfo{year}{2019}\natexlab{}.
\newblock \showarticletitle{Active acoustic switches using two-dimensional
  granular crystals}.
\newblock \bibinfo{journal}{\emph{Physical Review E}} \bibinfo{volume}{99},
  \bibinfo{number}{6} (\bibinfo{year}{2019}), \bibinfo{pages}{062901}.
\newblock


\bibitem[\protect\citeauthoryear{Yasuda, Buskohl, Gillman, Murphey, Stepney,
  Vaia, and Raney}{Yasuda et~al\mbox{.}}{2021}]%
        {yasuda2021mechanical}
\bibfield{author}{\bibinfo{person}{Hiromi Yasuda}, \bibinfo{person}{Philip~R
  Buskohl}, \bibinfo{person}{Andrew Gillman}, \bibinfo{person}{Todd~D Murphey},
  \bibinfo{person}{Susan Stepney}, \bibinfo{person}{Richard~A Vaia}, {and}
  \bibinfo{person}{Jordan~R Raney}.} \bibinfo{year}{2021}\natexlab{}.
\newblock \showarticletitle{Mechanical computing}.
\newblock \bibinfo{journal}{\emph{Nature}} \bibinfo{volume}{598},
  \bibinfo{number}{7879} (\bibinfo{year}{2021}), \bibinfo{pages}{39--48}.
\newblock


\bibitem[\protect\citeauthoryear{Zangeneh-Nejad, Sounas, Al{\`u}, and
  Fleury}{Zangeneh-Nejad et~al\mbox{.}}{2021}]%
        {zangeneh2021analogue}
\bibfield{author}{\bibinfo{person}{Farzad Zangeneh-Nejad},
  \bibinfo{person}{Dimitrios~L Sounas}, \bibinfo{person}{Andrea Al{\`u}}, {and}
  \bibinfo{person}{Romain Fleury}.} \bibinfo{year}{2021}\natexlab{}.
\newblock \showarticletitle{Analogue computing with metamaterials}.
\newblock \bibinfo{journal}{\emph{Nature Reviews Materials}}
  \bibinfo{volume}{6}, \bibinfo{number}{3} (\bibinfo{year}{2021}),
  \bibinfo{pages}{207--225}.
\newblock


\end{thebibliography}

\end{document}